# Generalized Public Key Transformations with Side Information

Gunjan Talati and Subhash Kak

**Abstract.** This paper presents results on generalized public key cryptography with exponentials modulo primes and composite numbers where the mapping is not one-to-one and the uniqueness is achieved by additional side information. Such transformations may be used for oblivious transfer and generate events of specific probabilities.

**Introduction**

This paper presents generalized public key transformations that require side information as extension to earlier proposals on cubic and quartic transformations [1],[2],[3]. Specifically, we present results on quintic and sextic transformations and then show how further generalizations may be made. For background papers on cryptographic protocols, the reader may see [4]-[7].

**The properties of the quintic transformation**

For quintic transformation $c = m^5 \bmod p$, five different values of message $m$ would give the same cipher $c$, where the value of prime $p$ is given by $p = 5k + 1$. The quintic roots of 1 would be 1, $\alpha$, $\alpha^2$, $\alpha^3$ and $\alpha^4$ and they are calculated by solving the equation

$$x^5 - 1 = 0 \qquad (1)$$

$$(x - 1)(x^4 + x^3 + x^2 + x^1 + 1) = 0 \qquad (2)$$

One of the roots is 1, and the other four roots are obtained by the following equation:

$$(x^4 + x^3 + x^2 + x^1 + 1) = 0$$

We convert the above quartic equation into depressed quartic equation by substituting

$$x = y - \frac{1}{4} \qquad (3)$$

and get $256y^4 + 160y^2 + 160y + 205 = 0$

Simplifying further,

$$\left(y^2 + \frac{5}{8}\right)^2 = \left(\frac{5}{8}y^2 - \frac{5}{8}y - \frac{105}{256}\right)$$

$$\left(y^2 + \frac{5}{8} + z\right)^2 = \left(\frac{5}{8} + 2z\right)y^2 - \frac{5}{8}y + \left(z^2 + \frac{5}{4}z - \frac{105}{256}\right) \qquad (4)$$

For the right hand side to be a perfect square,

$$\frac{25}{64} = 4\left(\frac{5}{8} + 2z\right)\left(z^2 + \frac{5}{4}z - \frac{105}{256}\right)$$

Solving the above equation, we get

$$4096z^3 + 6400z^2 - 80z - 725 = 0$$

The three values of z obtained are 0.31,-1.49,-.378. The first value $z = 0.31 = 5/16$. Substituting $z = 5/16$ in equation (4) we get

$$\left(y^2 + \frac{15}{16}\right)^2 = \left(\frac{\sqrt{5}}{2}y - \frac{\sqrt{5}}{8}\right)^2 \tag{5}$$

$$y = \frac{\frac{\sqrt{5}}{2} \pm \sqrt{\frac{-5 - \sqrt{5}}{2}}}{2}$$

and

$$y = \frac{-\frac{\sqrt{5}}{2} \pm \sqrt{\frac{-5 + \sqrt{5}}{2}}}{2}$$

$$x = \frac{\frac{\sqrt{5}}{2} \pm \sqrt{\frac{-5 - \sqrt{5}}{2}}}{2} - \frac{1}{4} \tag{6}$$

and

$$x = \frac{-\frac{\sqrt{5}}{2} \pm \sqrt{\frac{-5 + \sqrt{5}}{2}}}{2} - \frac{1}{4} \tag{7}$$

which can be rewritten as

$$\alpha_2 \text{ and } \alpha_3 = \frac{\frac{\sqrt{p+5}}{2} \pm \sqrt{\frac{-5 - \sqrt{p+5}}{2}}}{2} - \frac{1}{4} \tag{8}$$

$$\alpha_4 \text{ and } \alpha_5 = \frac{-\frac{\sqrt{p+5}}{2} \pm \sqrt{\frac{-5 + \sqrt{p+5}}{2}}}{2} - \frac{1}{4} \tag{9}$$

Alice sends $\alpha, n$ and $c$ to Bob. When Euler totient function $\varphi(n) = p - 1$ is not divisible by 25, Bob gets to know one of the five quintic roots by the inverse exponentiation equation:

$$c^{\frac{1}{5}} = c^{\frac{a(p-1)+5}{25}} \qquad (10)$$

where we pick the value of $a$ such that $\frac{a(p-1)+5}{25}$ comes out to be an integer.

The equations below show how equation (10) is derived.

$$c^{p-1} = 1 \qquad \text{(Fermat's little theorem)}$$

$$c^{a(p-1)} = 1$$

$$c^{a(p-1)}c^5 = c^5 \qquad \text{(multiplying } c^5 \text{ on both sides)}$$

$$c^{\frac{a(p-1)+5}{25}} = c^{\frac{1}{5}}$$

**Example 1.** Let $c = m^5 \bmod 61$. To obtain the five roots, we have to substitute the value of prime $p=61$ in equation (8) and (9).

We get the five roots as

$$\alpha_1 = 1$$

$$\alpha_2 = (441/4) = 34$$

$$\alpha_3 = (-391/4) = 9$$

$$\alpha_4 = (19/4) = 20$$

$$\alpha_5 = (-73/4) = 58$$

The exponent of message is 5 which is composed of only one prime 5. So, $\alpha^{\frac{5}{5}} \bmod 61 \neq 1$. Out of the five roots obtained above, $\alpha_1$ cannot be chosen as value of $\alpha$ because $1^1 \bmod 61 = 1$. Let $\alpha = 9$. Table 1 shows the 5-to-1 mapping for prime $p=61$ and $\alpha = 9$. Table 2 explains how the communication between Alice and Bob takes place.

**Table 1**: 5 to 1 Mapping for message $m$, prime $p=61$ and $\alpha=9$

| $m$ | $m\alpha$ | $m\alpha^2$ | $m\alpha^3$ | $m\alpha^4$ | $c = m^5 \bmod 61$ |
|---|---|---|---|---|---|
| 1 | 9 | 20 | 58 | 34 | 1 |
| 2 | 18 | 40 | 55 | 7 | 32 |
| 3 | 27 | 60 | 52 | 41 | 60 |
| 4 | 36 | 19 | 49 | 14 | 48 |
| 5 | 45 | 39 | 46 | 48 | 14 |
| 6 | 54 | 59 | 43 | 21 | 29 |
| 8 | 11 | 38 | 37 | 28 | 11 |
| 10 | 29 | 17 | 31 | 35 | 21 |
| 12 | 47 | 57 | 25 | 42 | 13 |
| 13 | 56 | 16 | 22 | 15 | 47 |
| 23 | 24 | 33 | 53 | 50 | 50 |
| 26 | 51 | 32 | 44 | 30 | 40 |

**Table 2**: Communication between Alice and Bob

| Alice | Bob |
|---|---|
| Let message $m$ chosen by Alice = 28.<br><br>After computing $m, m\alpha, m\alpha^2, m\alpha^3, m\alpha^4$ and arranging them in ascending order, Alice gets 8,11,28,37,38.<br><br>The rank of message $m$ = 28 is 3, so the side information based on the message chosen is 3.<br><br>Alice sends $c = m^5$ mod 61 = $28^5$ mod 61 = 11 and also sends side information=3 to Bob. | Using equation (10) Bob finds out that substituting a=2 gives the value of $\frac{a(p-1)+5}{25}$ as 5.<br><br>Bob gets one of the quintic roots as $11^5$ mod 61 = 11<br>and other roots as<br>(11*9) mod 61 = 38<br>(38*9) mod 61 = 37<br>(37*9) mod 61 = 28<br>(28*9) mod 61 = 8<br><br>Arranging them in ascending order, Bob gets 8,11,28,37,38 and side information=3 lets him know that the message chosen by Alice was $m$ = 28. |

**Quintic transformation modulo a composite number**

For composite number, $n$ is chosen such that it is composed of two primes given by $n = pq$.

We will show examples where the Euler totient function $\varphi(n) = (p-1)(q-1)$, is divisible by 5 but not by 25.

Inverse exponentiation operation for quintic transformation modulo composite number is calculated by

$$c^{\frac{1}{5}} = c^{\frac{a\varphi(n)+5}{25}} \qquad (11)$$

where we pick the value of $a$ such that $\frac{a\varphi(n)+5}{25}$ comes out to be an integer.

**Example 2.** Let $c = m^5$ mod 187, $p$ =11, $q$ =17. The quintic roots for $p$ obtained by substituting $p$=11 in equation (8),(9) are $\alpha_p$ = 1, 4, 5, 9, 3

Since we already know the five quintic roots of $p$, we only need one quintic root of $q$ to apply CRT.
One of the root for quintic transformation will always be 1, therefore we select $\alpha_q$=1.

$\alpha_q$=1

$\alpha = (\alpha_p \times q \times ||q||_p^{-1} + \alpha_q \times p \times ||p||_q^{-1}) \mod pq$

$\alpha_1$ = ( (1x17x2)+(1x11x14) ) mod 187 = 188 mod 187 = 1

$\alpha_2$ = ( (4x17x2)+(1x11x14) ) mod 187 = 290 mod 187 = 103

$\alpha_3$ = ( (5x17x2)+(1x11x14) ) mod 187 = 324 mod 187 = 137

$\alpha_4$ = ( (9x17x2)+(1x11x14) ) mod 187 = 460 mod 187 = 86

$$\alpha_5 = (\ (3 \times 17 \times 2) + (1 \times 11 \times 14)\ )\ \text{mod}\ 187 = 256\ \text{mod}\ 187 = 69$$

Any of the five roots obtained other than 1 can be chosen as value of $\alpha$. Let $\alpha$ = 69. Table 3 explains how the communication between Alice and Bob takes place.

**Table 3**: Communication between Alice and Bob

| Alice | Bob |
|---|---|
| Let message $m$ chosen by Alice = 3.<br><br>After computing $m, m\alpha, m\alpha^2, m\alpha^3, m\alpha^4$ and arranging them in ascending order, Alice gets 3,20,37,71,122.<br><br>The rank of message $m$ = 3 is 1, so the side information based on the message chosen is 1.<br><br>Alice sends $c = m^5 \text{mod}\ 187 = 3^5\ \text{mod}\ 187 = 56$ and also sends side information=1 to Bob. | Using equation (11) Bob finds out that substituting a=2 gives the value of $\frac{a(p-1)(q-1)+5}{25}$ as 13.<br><br>Bob gets one of the quintic roots as $56^{13}$ mod 187 = 122<br>and other roots as<br>(122*69) mod 187 = 3<br>(3*69) mod 187 = 20<br>(20*69) mod 187 = 71<br>(71*69) mod 187 = 37<br><br>Arranging them in ascending order, Bob gets 3,20,37,71,122 and side information=1 lets him know that the message chosen by Alice was $m$ = 3. |

**φ(n) divisible by 25**

When φ(n) is divisible by 25, the 25 quintic roots of 1 can be obtained from the equation

$$\alpha^{25} - 1 = 0 \qquad (12)$$

$$(\alpha^5)^5 - 1 = 0$$

$$(\alpha^5 - 1)(\alpha^{20} + \alpha^{15} + \alpha^{10} + \alpha^5 + 1) = 0 \qquad (13)$$

The message that Alice wants to send to Bob would be multiplied with 25 roots of 1 and arranged in ascending order so that its position number in the set of 25 could be find out. After that, Alice sends $c = m^5 \text{mod}\ n$ along with the position number of message as side information.

Bob will obtain $c$. He will find the quintic root of $c$, let's say its x. Bob would then find the multiple of x with all the roots of 1, arrange them in ascending order and the position number sent as side information lets him know about the message $m$ that was sent by Alice.

**Example 3.** Let $c = m^5 \text{mod}\ 341$, $p$ =31, $q$ =11. The quintic roots for $p$ obtained by substituting $p$=31 in equation (8),(9) are $\alpha_p$ = 1, 2, 4, 8, 16. The quintic roots for $q$ obtained by substituting $q$=11 in equation (8),(9) are $\alpha_q$= 1, 3, 4, 5, 9

By using Chinese Remainder Theorem (CRT), we get the 25 quintic roots of 1 as follows:

$$\alpha = (\ \alpha_p\ \text{x}\ q\ \text{x}\ ||q||_p^{-1} + \alpha_q\ \text{x}\ p\ \text{x}\ ||p||_q^{-1}\ )\ \text{mod}\ pq \qquad (14)$$

$\alpha_1 = ( (1 \times 11 \times 17) + (1 \times 31 \times 5) ) \bmod 341 = 342 \bmod 341 = 1$

$\alpha_2 = ( (2 \times 11 \times 17) + (1 \times 31 \times 5) ) \bmod 341 = 529 \bmod 341 = 188$

$\alpha_3 = ( (4 \times 11 \times 17) + (1 \times 31 \times 5) ) \bmod 341 = 903 \bmod 341 = 221$

$\alpha_4 = ( (8 \times 11 \times 17) + (1 \times 31 \times 5) ) \bmod 341 = 1651 \bmod 341 = 287$

$\alpha_5 = ( (16 \times 11 \times 17) + (1 \times 31 \times 5) ) \bmod 341 = 3147 \bmod 341 = 78$

$\alpha_6 = ( (1 \times 11 \times 17) + (3 \times 31 \times 5) ) \bmod 341 = 652 \bmod 341 = 311$

$\alpha_7 = ( (2 \times 11 \times 17) + (3 \times 31 \times 5) ) \bmod 341 = 839 \bmod 341 = 157$

$\alpha_8 = ( (4 \times 11 \times 17) + (3 \times 31 \times 5) ) \bmod 341 = 1213 \bmod 341 = 190$

$\alpha_9 = ( (8 \times 11 \times 17) + (3 \times 31 \times 5) ) \bmod 341 = 1961 \bmod 341 = 256$

$\alpha_{10} = ( (16 \times 11 \times 17) + (3 \times 31 \times 5) ) \bmod 341 = 3457 \bmod 341 = 47$

$\alpha_{11} = ( (1 \times 11 \times 17) + (4 \times 31 \times 5) ) \bmod 341 = 807 \bmod 341 = 125$

$\alpha_{12} = ( (2 \times 11 \times 17) + (4 \times 31 \times 5) ) \bmod 341 = 994 \bmod 341 = 312$

$\alpha_{13} = ( (4 \times 11 \times 17) + (4 \times 31 \times 5) ) \bmod 341 = 1368 \bmod 341 = 4$

$\alpha_{14} = ( (8 \times 11 \times 17) + (4 \times 31 \times 5) ) \bmod 341 = 2116 \bmod 341 = 70$

$\alpha_{15} = ( (16 \times 11 \times 17) + (4 \times 31 \times 5) ) \bmod 341 = 3612 \bmod 341 = 202$

$\alpha_{16} = ( (1 \times 11 \times 17) + (5 \times 31 \times 5) ) \bmod 341 = 962 \bmod 341 = 280$

$\alpha_{17} = ( (2 \times 11 \times 17) + (5 \times 31 \times 5) ) \bmod 341 = 1149 \bmod 341 = 126$

$\alpha_{18} = ( (4 \times 11 \times 17) + (5 \times 31 \times 5) ) \bmod 341 = 1523 \bmod 341 = 159$

$\alpha_{19} = ( (8 \times 11 \times 17) + (5 \times 31 \times 5) ) \bmod 341 = 2271 \bmod 341 = 225$

$\alpha_{20} = ( (16 \times 11 \times 17) + (5 \times 31 \times 5) ) \bmod 341 = 3767 \bmod 341 = 16$

$\alpha_{21} = ( (1 \times 11 \times 17) + (9 \times 31 \times 5) ) \bmod 341 = 1582 \bmod 341 = 218$

$\alpha_{22} = ( (2 \times 11 \times 17) + (9 \times 31 \times 5) ) \bmod 341 = 1769 \bmod 341 = 64$

$\alpha_{23} = ( (4 \times 11 \times 17) + (9 \times 31 \times 5) ) \bmod 341 = 2143 \bmod 341 = 97$

$\alpha_{24} = ( (8 \times 11 \times 17) + (9 \times 31 \times 5) ) \bmod 341 = 2891 \bmod 341 = 163$

$\alpha_{25} = ( (16 \times 11 \times 17) + (9 \times 31 \times 5) ) \bmod 341 = 4387 \bmod 341 = 295$

Hence, the 25 roots of 1 obtained are {1, 188, 221, 287, 78, 311, 157, 190, 256, 47, 125, 312, 4, 70, 202, 280, 126, 159, 225, 16, 218, 64, 97, 163, 295}. Let message $m$ that Alice wants to send to Bob is 51. Multiplying 51 with 25 roots of 1 and arranging them in ascending order, Alice obtains {10, 18, 40, 41, 51, 72, 98, 129, 134, 142, 160, 164, 173, 175, 195, 204, 206, 222, 226, 227, 237, 266, 288, 299, 315}. The position number of 51 in the set is 5. Alice sends $c = m^5 \bmod$

341 = 87 along with side information 5 to Bob. Bob gets the cipher and calculates quintic root of 87 by using CRT which comes out to be 222. Multiplying 222 with 25 roots of 1 and arranging them in ascending order, Bob obtains {10, 18, 40, 41, 51, 72, 98, 129, 134, 142, 160,164, 173, 175, 195, 204, 206, 222, 226, 227, 237, 266, 288, 299, 315}. The side information 5 lets him know that the message sent by Alice was $m=51$.

**Probability events**

The 25 quintic roots of 1 can be put down in a different way as a number, its square, its cube and fourth degree.

$$1, a, a^2, a^3, a^4, b, b^2, b^3, b^4, c, c^2, c^3, c^4, d, d^2, d^3, d^4, e, e^2, e^3, e^4, f, f^2, f^3, f^4 \quad (15)$$

They can be put into six different groups as follows:

$1, a, a^2, a^3, a^4;$  $\quad 1, b, b^2, b^3, b^4;$  $\quad 1, c, c^2, c^3, c^4;$  $\quad 1, d, d^2, d^3, d^4;$
$1, e, e^2, e^3, e^4;$  $\quad 1, f, f^2, f^3, f^4$

For the example above, we get six different groups as:

{1, 4, 16, 64, 256}, {1, 47, 163, 159, 312}, {1, 70, 126, 295, 190}, {1, 78, 287, 221, 188}, {1, 97, 157, 202, 225}, {1, 125, 280, 218, 311}

**The properties of the sextic transformation**

For sextic transformation $c = m^6$ mod $p$, six different values of message $m$ would give the same cipher $c$, where the value of prime $p$ is given by $p = 6k + 1$.
We will show example where the Euler totient function $\varphi(n)$, is divisible by 6 but not by 36.

The sextic roots of 1 would be 1, $\alpha, \alpha^2, \alpha^3, \alpha^4$ and $\alpha^5$ and they are calculated by solving the equation

$$\alpha^6 - 1 = 0 \quad (16)$$

$(\alpha - 1)(\alpha^5 + \alpha^4 + \alpha^3 + \alpha^2 + \alpha + 1) = 0$

$(\alpha - 1)(\alpha + 1)(\alpha^4 + \alpha^2 + 1) = 0$

$(\alpha - 1)(\alpha + 1)((\alpha^2)^2 + \alpha^2 + 1) = 0 \quad (17)$

Six roots obtained from the above equation are:

$\alpha_1 = 1 \quad (18)$

$\alpha_2 = -1 = (p - 1) \quad (19)$

$$\alpha_3 = \sqrt{\frac{-1 + \sqrt{-3}}{2}} = \sqrt{\frac{-1 + \sqrt{p - 3}}{2}} \quad (20)$$

$$\alpha_4 = -\sqrt{\frac{-1+\sqrt{-3}}{2}} = -\sqrt{\frac{-1+\sqrt{p-3}}{2}} \qquad (21)$$

$$\alpha_5 = \sqrt{\frac{-1-\sqrt{-3}}{2}} = \sqrt{\frac{-1-\sqrt{p-3}}{2}} \qquad (22)$$

$$\alpha_6 = -\sqrt{\frac{-1-\sqrt{-3}}{2}} = -\sqrt{\frac{-1-\sqrt{p-3}}{2}} \qquad (23)$$

Alice sends $\alpha$, $n$ and $c$ to Bob. When Euler totient function $\varphi(n) = p - 1$ is not divisible by 36, Bob gets to know one of the six sextic roots by the inverse exponentiation equation:

$$c^{\frac{1}{6}} = c^{\frac{a(p-1)+6}{36}} \qquad (24)$$

where we pick the value of $a$ such that $\frac{a(p-1)+6}{36}$ comes out to be an integer.

**Example 4.** Let $c = m^6 \bmod 43$. To obtain the six roots, we have to substitute the value of prime $p=43$ in equations (19) to (23) which provide the roots in addition to 1. We get the six roots as

$\alpha_1 = 1$

$\alpha_2 = -1 = 42$

$$\alpha_3 = \sqrt{\frac{-1+\sqrt{-3}}{2}} = \sqrt{\frac{-1+\sqrt{(43*4)-3}}{2}} = \sqrt{6} = 7$$

$$\alpha_4 = -\sqrt{\frac{-1+\sqrt{-3}}{2}} = -\sqrt{\frac{-1+\sqrt{(43*4)-3}}{2}} = -\sqrt{6} = -7 = 36$$

$$\alpha_5 = \sqrt{\frac{-1-\sqrt{-3}}{2}} = \sqrt{\frac{-1-\sqrt{(43*4)-3}}{2}} = \sqrt{-7} = 6$$

$$\alpha_6 = -\sqrt{\frac{-1-\sqrt{-3}}{2}} = -\sqrt{\frac{-1-\sqrt{(43*4)-3}}{2}} = -\sqrt{-7} = -6 = 37$$

The exponent of message is 6 which is composed of two primes 2 and 3. So, $\alpha^{\frac{6}{2}} \bmod 43 \neq 1$ and $\alpha^{\frac{6}{3}} \bmod 43 \neq 1$. Out of the six roots obtained above, 1 and 42 cannot be chosen as value of $\alpha$ because $1^2 \bmod 43 = 1$ and $42^2 \bmod 43 = 1$. Also, 6 and 36 cannot be chosen as value of $\alpha$ because $6^3 \bmod 43 = 1$ and $36^3 \bmod 43 = 1$ Therefore, $\alpha$ can be either $\alpha_3$ or $\alpha_6$. Let $\alpha = 7$.

Table 4 shows the 6-to-1 mapping for prime $p=43$ and $\alpha = 7$. Table 5 explains how the communication between Alice and Bob takes place.

**Table 4**: 6 to 1 Mapping for message $m$, prime $p=43$ and $\alpha=7$

| $m$ | $m\alpha$ | $m\alpha^2$ | $m\alpha^3$ | $m\alpha^4$ | $m\alpha^5$ | $c = m^6 \bmod 43$ |
|---|---|---|---|---|---|---|
| 1 | 7 | 6 | 42 | 36 | 37 | 1 |
| 2 | 14 | 12 | 41 | 29 | 31 | 21 |
| 3 | 21 | 18 | 40 | 22 | 25 | 41 |
| 4 | 28 | 24 | 39 | 15 | 19 | 11 |
| 5 | 35 | 30 | 38 | 8 | 13 | 16 |
| 9 | 20 | 11 | 34 | 23 | 32 | 4 |
| 10 | 27 | 17 | 33 | 16 | 26 | 35 |

**Table 5**: Communication between Alice and Bob

| Alice | Bob |
|---|---|
| Let message $m$ chosen by Alice = 2.<br><br>After computing $m, m\alpha, m\alpha^2, m\alpha^3, m\alpha^4, m\alpha^5$ and arranging them in ascending order, Alice gets 2,12,14,29,31,41.<br><br>The rank of message $m = 2$ is 1, so the side information based on the message chosen is 1.<br><br>Alice sends $c = m^6 \bmod 43 = 2^6 \bmod 43 = 21$ and also sends side information=1 to Bob. | Using equation (24) Bob finds out that substituting a=5 gives the value of $\frac{a(p-1)+6}{36}$ as 6.<br><br>Bob gets one of the sextic roots as $21^6 \bmod 43 = 41$<br>and other roots as<br>(41*7) mod 43 = 29<br>(29*7) mod 43 = 31<br>(31*7) mod 43 = 2<br>(2*7) mod 43 = 14<br>(14*7) mod 43 = 12<br><br>Arranging them in ascending order, Bob gets 2,12,14,29,31,41 and side information=1 lets him know that the message chosen by Alice was $m = 2$. |

**Sextic transformation modulo a composite number**

For composite number, $n$ is chosen such that it is composed of two primes given by $n = pq$. The inverse exponentiation operation for sextic transformation modulo composite number is calculated by

$$c^{\frac{1}{6}} = c^{\frac{a\varphi(n)+6}{36}} \tag{25}$$

where we pick the value of $a$ such that $\frac{a\varphi(n)+6}{36}$ comes out to be an integer.

**φ(n) divisible by 36**

When $\varphi(n)$ is divisible by 36, the 36 sextic roots of 1 can be obtained from the equation

$$\alpha^{36} - 1 = 0 \tag{26}$$

$$(\alpha^6)^6 - 1 = 0$$

$$(\alpha^6 - 1)(\alpha^6 + 1)((\alpha^{12})^2 + \alpha^{12} + 1) = 0 \tag{27}$$

The message that Alice wants to send to Bob would be multiplied with 36 roots of 1 and arranged in ascending order so that its position number in the set of 36 could be find out. After that, Alice sends $c = m^6 \mod n$ along with the position number of message as side information.

Bob will obtain $c$. He will find the sextic root of $c$, let's say its x. Bob would then find the multiple of x with all the roots of 1, arrange them in ascending order and the position number sent as side information lets him know about the message $m$ that was sent by Alice.

**Example 5.** Let $c = m^6 \mod 403$, $p = 31$, $q = 13$. The sextic roots for $p$ are obtained by substituting $p=31$ in equation (19) to (23) and we get $\alpha_p$ = 1, 5, 6, 25, 26, 30. Likewise the sextic roots for $q$ are obtained by substituting $q=13$ and we finally obtain $\alpha_q$ = 1, 3, 4, 9, 10, 12.

By using Chinese Remainder Theorem (CRT), we get the 36 sextic roots of 1 as follows:

$$\alpha = (\alpha_p \times q \times ||q||_p^{-1} + \alpha_q \times p \times ||p||_q^{-1}) \mod pq \tag{28}$$

$\alpha_1 = ((1 \times 13 \times 12)+(1 \times 31 \times 8)) \mod 403 = 404 \mod 403 = 1$

$\alpha_2 = ((5 \times 13 \times 12)+(1 \times 31 \times 8)) \mod 403 = 1028 \mod 403 = 222$

$\alpha_3 = ((6 \times 13 \times 12)+(1 \times 31 \times 8)) \mod 403 = 1184 \mod 403 = 378$

$\alpha_4 = ((25 \times 13 \times 12)+(1 \times 31 \times 8)) \mod 403 = 4148 \mod 403 = 118$

$\alpha_5 = ((26 \times 13 \times 12)+(1 \times 31 \times 8)) \mod 403 = 4304 \mod 403 = 274$

$\alpha_6 = ((30 \times 13 \times 12)+(1 \times 31 \times 8)) \mod 403 = 4928 \mod 403 = 92$

$\alpha_7 = ((1 \times 13 \times 12)+(3 \times 31 \times 8)) \mod 403 = 900 \mod 403 = 94$

$\alpha_8 = ((5 \times 13 \times 12)+(3 \times 31 \times 8)) \mod 403 = 1524 \mod 403 = 315$

$\alpha_9 = ((6 \times 13 \times 12)+(3 \times 31 \times 8)) \mod 403 = 1680 \mod 403 = 68$

$\alpha_{10} = ((25 \times 13 \times 12)+(3 \times 31 \times 8)) \mod 403 = 4644 \mod 403 = 211$

$\alpha_{11} = ((26 \times 13 \times 12)+(3 \times 31 \times 8)) \mod 403 = 4800 \mod 403 = 367$

$\alpha_{12} = ((30 \times 13 \times 12)+(3 \times 31 \times 8)) \mod 403 = 5424 \mod 403 = 185$

$\alpha_{13} = ((1 \times 13 \times 12)+(4 \times 31 \times 8)) \mod 403 = 1148 \mod 403 = 342$

$\alpha_{14} = ((5 \times 13 \times 12)+(4 \times 31 \times 8)) \mod 403 = 1772 \mod 403 = 160$

$\alpha_{15} = ((6 \times 13 \times 12)+(4 \times 31 \times 8)) \mod 403 = 1928 \mod 403 = 316$

$\alpha_{16} = ((25 \times 13 \times 12)+(4 \times 31 \times 8)) \mod 403 = 4892 \mod 403 = 56$

$\alpha_{17} = ((26 \times 13 \times 12)+(4 \times 31 \times 8)) \mod 403 = 5048 \mod 403 = 212$

$\alpha_{18} = ((30 \times 13 \times 12)+(4 \times 31 \times 8)) \mod 403 = 5672 \mod 403 = 30$

$\alpha_{19}$ = ( (1x13x12)+(9x31x8) ) mod 403 = 2388 mod 403 = 373

$\alpha_{20}$ = ( (5x13x12)+(9x31x8) ) mod 403 = 3012 mod 403 = 191

$\alpha_{21}$ = ( (6x13x12)+(9x31x8) ) mod 403 = 3168 mod 403 = 347

$\alpha_{22}$ = ( (25x13x12)+(9x31x8) ) mod 403 = 6132 mod 403 = 87

$\alpha_{23}$ = ( (26x13x12)+(9x31x8) ) mod 403 = 6288 mod 403 = 243

$\alpha_{24}$ = ( (30x13x12)+(9x31x8) ) mod 403 = 6912 mod 403 = 61

$\alpha_{25}$ = ( (1x13x12)+(10x31x8) ) mod 403 = 2636 mod 403 = 218

$\alpha_{26}$ = ( (5x13x12)+(10x31x8) ) mod 403 = 3260 mod 403 = 36

$\alpha_{27}$ = ( (6x13x12)+(10x31x8) ) mod 403 = 3416 mod 403 = 192

$\alpha_{28}$ = ( (25x13x12)+(10x31x8) ) mod 403 = 6380 mod 403 = 335

$\alpha_{29}$ = ( (26x13x12)+(10x31x8) ) mod 403 = 6536 mod 403 = 88

$\alpha_{30}$ = ( (30x13x12)+(10x31x8) ) mod 403 = 7160 mod 403 = 309

$\alpha_{31}$ = ( (1x13x12)+(12x31x8) ) mod 403 = 3132 mod 403 = 311

$\alpha_{32}$ = ( (5x13x12)+(12x31x8) ) mod 403 = 3756 mod 403 = 129

$\alpha_{33}$ = ( (6x13x12)+(12x31x8) ) mod 403 = 3912 mod 403 = 285

$\alpha_{34}$ = ( (25x13x12)+(12x31x8) ) mod 403 = 6876 mod 403 = 25

$\alpha_{35}$ = ( (26x13x12)+(12x31x8) ) mod 403 = 7032 mod 403 = 181

$\alpha_{36}$ = ( (30x13x12)+(12x31x8) ) mod 403 = 7656 mod 403 = 402

Hence, the 36 roots of 1 obtained are {1, 222, 378, 118, 274, 92, 94, 315, 68, 211, 367, 185, 342, 160, 316, 56, 212, 30, 373, 191, 347, 87, 243, 61, 218, 36, 192, 335, 88, 309, 311, 129, 285, 25, 181, 402}. Let message $m$ that Alice wants to send to Bob is 59. Multiplying 59 with 36 roots of 1 and arranging them in ascending order, Alice obtains {15,18,28,34,44,46,47,59,80,96,106,109,111,137,158,171,189,201,202,214,232,245,266,292,294,297,307,323,344,356,357,359,369,375,385,388}. The position number of 59 in the set is 8. Alice sends $c = m^6$ mod 403 = 233 along with side information 8 to Bob. Bob gets the cipher and calculates sextic root of 233 by using CRT which comes out to be 158. Multiplying 158 with 36 roots of 1 and arranging them in ascending order, Bob obtains {15,18,28,34,44,46,47,59,80,96,106,109,111,137,158,171,189,201,202,214,232,245,266,292,294,297,307,323,344,356,357,359,369,375,385,388}. The side information 8 lets him know that the message sent by Alice was $m$=59.

**Probability events**

The 36 sextic roots of 1 can be put down in a different way as a number, its square, its cube, fourth degree and fifth degree.

$1, a, a^2, a^3, a^4, a^5, b, b^2, b^3, b^4, b^5, c, c^2, c^3, c^4, c^5, d, d^2, d^3, d^4, d^5, e, e^2, e^3, e^4, e^5, f, f^2, f^3, f^4, f^5, g, g^2, g^3, g^4, g^5, h, h^2, h^3, h^4, h^5, i, i^2, i^3, i^4, i^5, j, j^2, j^3, j^4, j^5, k, k^2, k^3, k^4, k^5, l, l^2, l^3, l^4, l^5$

Out of the 61 values shown above 8 values would be repeated three times and 3 values would be repeated four times which would make 36 sextic roots of 1.

They can be put into twelve different groups as follows:

$1, a, a^2, a^3, a^4, a^5$;   $1, b, b^2, b^3, b^4, b^5$;   $1, c, c^2, c^3, c^4, c^5$;   $1, d, d^2, d^3, d^4, d^5$;
$1, e, e^2, e^3, e^4, e^5$;   $1, f, f^2, f^3, f^4, f^5$;   $1, g, g^2, g^3, g^4, g^5$;   $1, h, h^2, h^3, h^4, h^5$;
$1, i, i^2, i^3, i^4, i^5$;   $1, j, j^2, j^3, j^4, j^5$;   $1, k, k^2, k^3, k^4, k^5$;   $1, l, l^2, l^3, l^4, l^5$;

If $x = a, b, c, d, e, f, g, h, i, j, k, l$, then $x^2 \bmod n \neq 1$ and $x^3 \bmod n \neq 1$. If $x^2 \bmod n = 1$, then it would generate groups as $\{1, x, 1, x, 1, x\}$. If $x^3 \bmod n = 1$, then it would generate groups as $\{1, x, x^2, 1, x, x^2\}$

For the example above, we get twelve different groups as:

{1, 25, **222**, *311*, **118**, 129}, {1, 30, **94**, *402*, **373**, 309}, {1, 36, **87**, *311*, **315**, 56},
{1, 61, **94**, *92*, **373**, 185}, {1, 68, **191**, *92*, **211**, 243}, {1, 88, **87**, *402*, **315**, 316},
{1, 160, **211**, *311*, **191**, 335}, {1, 181, **118**, *402*, **222**, 285}, {1, 192, **191**, *402*, **211**, 212},
{1, 218, **373**, *311*, **94**, 342}, {1, 274, **118**, *92*, **222**, 378}, {1, 367, **87**, *92*, **315**, 347}

The 8 values which are repeated three times are 222,94,211,373,191,118,87 and 315. The 3 values which are repeated four times are 311,92 and 402.

## Transformation modulo a prime number

Table 6 shows how the communication between Alice and Bob takes place where $t$ is the exponent used to generate the cipher $c = m^t \bmod p$ for a message $m$ and $p$ is a prime number given by $p = tk + 1$

## Transformation where $\varphi(n)$ is divisible by $t$ but not by $t^2$

Step 1: When one of the prime is of the form $tk + 1$, then $t$ roots of 1 for that prime can be combined with one obvious root 1 of other prime to obtain $t$ roots of 1 for composite $n$ by using the CRT equation:

$$\alpha = ((\alpha_p \text{ x } q \text{ x } a) + (\alpha_q \text{ x } p \text{ x } b)) \bmod pq \qquad (28)$$

In the above equation $\alpha_p$ and $\alpha_q$ are roots of prime $p$ and $q$ respectively and $a$ and $b$ are values that satisfy the equation $qa \bmod p = 1$ and $pb \bmod q = 1$ respectively. The remaining steps are same as Table 6 except prime p gets substituted by composite n given by n=pq and φ(p) gets substituted by φ(n) which is given by (p-1)(q-1).

| Alice | Bob |
|---|---|
| **Step 1:**<br>In Transformation modulo a Prime number, Prime $p$ is selected such that Euler totient function $\varphi(p) = (p-1)$ is divisible by $t$ but not by $t^2$.<br>Alice finds out the $t$ roots of 1 for prime $p$ by the equation $\alpha^t - 1 = 0$. Let's call this roots as $\alpha_1, \alpha_2 \dots \alpha_t$. Since the equation is of the form $\alpha^t - 1 = 0$, $\alpha_1$ will always be 1.<br>Alternative method is to select $m$ from 1 to $p - 1$, and those values of $m$ which gives $m^t \bmod p = 1$ are the roots of 1.<br><br>**Step 2:**<br>Select $\alpha$ from the set $\{\alpha_1, \alpha_2 \dots \alpha_t\}$.<br>Ignore those values of $\alpha$ which gives repeating values in $\{1, \alpha, \alpha^2 \dots \alpha^{t-1}\} \bmod p$.<br><br>**Step 3:**<br>Create a table of $\frac{\varphi(p)}{t}$ rows and $(t+1)$ columns where the first $t$ column represents $m, m\alpha, m\alpha^2 \dots m\alpha^{t-1}$ and the last column represents $c = m^t \bmod p$. Select $m$ from $\{1,2,3 \dots p-1\}$ and fill the table in such a way that $\frac{\varphi(p)}{t}$ rows and $t$ columns displays $\varphi(p)$ unique values. For every row, first $t$ column represents $t$ different message that generates the same cipher given by the last column. Hence, $t$-to-1 mapping is observed here.<br><br>**Step 4:**<br>Alice computes $m, m\alpha, m\alpha^2 \dots m\alpha^{t-1}$. After arranging them in ascending order, she finds out the position of her message in the set of ascending order. The position is sent as side information to Bob along with the cipher $c = m^t \bmod p$. | **Step 5:**<br>When Euler totient function $\varphi(p) = (p-1)$ is not divisible by $t^2$, Bob gets to know one of the $t$ roots by the inverse exponentiation equation:<br>$$c^{\frac{1}{t}} = c^{\frac{a\varphi(p)+t}{t^2}}$$<br>where substituting an integer value of $a$ gives the exponent on the R.H.S. $\frac{a\varphi(p)+t}{t^2}$ as an integer. Let's call the resulting value $\frac{a\varphi(p)+t}{t^2}$ as $res$.<br><br>**Step 6:**<br>Bob gets one of the $t$ roots as $\alpha_1 = c^{res} \bmod p$ and other roots as<br>$(\alpha_1 * \alpha) \bmod p = \alpha_2$<br>$(\alpha_2 * \alpha) \bmod p = \alpha_3$<br>……<br>……<br>$(\alpha_{t-1} * \alpha) \bmod p = \alpha_t$<br><br>**Step 7:**<br>Arranging $\{\alpha_1, \alpha_2 \dots \alpha_t\}$ in ascending order and using side information sent by Alice specifying the position of message in the set of ascending order, Bob gets to know about the message $m$ chosen by Alice. |

**Table 6:** Communication between Alice and Bob

## Transformation where $\varphi(n)$ is divisible by both $t$ and $t^2$

Step 1: When both the primes are of the form $tk + 1$, then $t$ roots of 1 for one prime can be combined with $t$ roots of 1 for other prime to obtain $t^2$ roots of 1 for composite $n$ by the use of CRT.

Step 2: The message that Alice wants to send to Bob would be multiplied with $t^2$ roots of 1 and arranged in ascending order so that position of message can be sent as side information. Alice sends cipher $c = m^t \bmod n$ along with the side information to Bob.

Step 3: Bob finds out the inverse of the cipher by using CRT. Let's say its $x$. Bob would then multiply $x$ with $t^2$ roots of 1 and arrange the set obtained in ascending order. The side information lets him know about the message $m$ that was sent by Alice.

**Probability events when $\varphi(n)$ is divisible by both $t$ and $t^2$**

The $t^2$ roots of 1 may be put down in a different way as a number, its square, its cube and so on and different number of groups can be formed. If Alice and Bob agree that each of these groups begins with a 1, then they can be put into $g$ different groups as follows:

$$1, a, a^2, a^3 \ldots a^{t-1}; \quad 1, b, b^2, b^3 \ldots b^{t-1}; \quad 1, c, c^2, c^3 \ldots c^{t-1}; \quad \ldots \ldots \ldots \ldots \ldots \ldots \ldots \ldots \ldots 1, i, i^2, i^3 \ldots i^{t-1} \quad (29)$$

The number of groups $g$ obtained would be:

$$g = \frac{t^2 - (t - \varphi(t))^2}{\varphi(t)} \quad \text{or} \quad g = \frac{t^2 - \left(\frac{t}{a}\right)^2}{\varphi(t)}$$

when the exponent of message is composed of only one prime $a$.

$$g = \frac{t^2 - \left[\left(\frac{t}{a}\right)^2 + \left(\frac{t}{b}\right)^2 - \left[\gcd\left(\frac{t}{a}, \frac{t}{b}\right)\right]^2\right]}{\varphi(t)}$$

when the exponent of message is composed of exactly two primes $a$ and $b$. The $i$ in equation (29) represents the character corresponding to the number of groups $g$ obtained.

Out of the the $t^2$ roots of 1 obtained for composite $n$, some of the roots cannot be selected as initial value $\{a, b, c \ldots \ldots i\}$ as they gives repeating values in the group which they correspond to. There are two cases for finding out the roots which cannot be selected as initial value.

**Case 1: When the exponent of message is composed of only one prime $a$.**

Calculate $t - \varphi(t)$ or $\frac{t}{a}$. Let $t - \varphi(t) = x$ or $\frac{t}{a} = x$. Now, out of the $t$ roots obtained for each of the prime $p$ and $q$ of composite $n$, we need to find 1 root $\alpha_1$ of prime $p$ and 1 root $\alpha_2$ of prime $q$ such that

$$(1 + \alpha_1 + \alpha_1^2 \ldots \ldots \alpha_1^{x-1}) \bmod p = 0 \text{ and}$$

$$(1 + \alpha_2 + \alpha_2^2 \ldots \ldots \alpha_2^{x-1}) \bmod q = 0$$

The $x$ roots $(1, \alpha_1, \alpha_1^2 \ldots \ldots \alpha_1^{x-1})$ of prime $p$ and $x$ roots $(1, \alpha_2, \alpha_2^2 \ldots \ldots \alpha_2^{x-1})$ of prime $q$ when combined using CRT gives a total of $x^2$ values. These are the $x^2$ values which cannot be selected as initial value.

**Case 2: When the exponent of message is composed of exactly two primes $a$ and $b$**

Calculate $\frac{t}{a}$ and $\frac{t}{b}$. Let $\frac{t}{a} = x$ and $\frac{t}{b} = y$. Now, out of the $t$ roots obtained for each of the prime $p$ and $q$ of composite $n$, we need to find two roots $\alpha_1, \alpha_2$ of prime $p$ and two roots $\alpha_3, \alpha_4$ of prime $q$ such that

$$(1 + \alpha_1 + \alpha_1^2 \ldots \ldots \alpha_1^{x-1}) \bmod p = 0$$

$$(1 + \alpha_2 + \alpha_2{}^2 \ldots \alpha_2{}^{y-1}) \bmod p = 0 \text{ and}$$

$$(1 + \alpha_3 + \alpha_3{}^2 \ldots \alpha_3{}^{x-1}) \bmod q = 0$$

$$(1 + \alpha_4 + \alpha_4{}^2 \ldots \alpha_4{}^{y-1}) \bmod q = 0$$

$x$ roots $(1, \alpha_1, \alpha_1{}^2 \ldots \alpha_1{}^{x-1})$ of prime $p$ and $x$ roots $(1, \alpha_3, \alpha_3{}^2 \ldots \alpha_3{}^{x-1})$ of prime $q$ when combined using CRT gives a total of $x^2$ values which cannot be selected as initial value. Similarly, $y$ roots $(1, \alpha_2, \alpha_2{}^2 \ldots \alpha_2{}^{y-1})$ of prime $p$ and $y$ roots $(1, \alpha_4, \alpha_4{}^2 \ldots \alpha_4{}^{y-1})$ of prime $q$ when combined using CRT gives a total of $y^2$ values which cannot be selected as initial value.

If the numbers of groups $g$ formed are considered as a matrix of $g$ rows and $t$ columns having index of rows and columns starting from 0, then the values which cannot be selected as initial value can be found in the columns having index of column not relatively prime to $t$.

**Conclusions**

We have shown how the many-to-one mapping for the exponents 5 and 6 can be used for public key cryptography if additional side information is used. These mappings can also be used for oblivious transfer or multiparty communications.